\documentclass[aps,preprint,amsmath,amssymb]{revtex4}
\usepackage{graphicx}
\usepackage{textcomp}
\usepackage{times}
\usepackage{color}
\usepackage{amsmath,amsfonts}

\usepackage{booktabs}
\usepackage{threeparttable}
\usepackage{epsfig}
\begin{document}

\begin{flushleft}
{\Large
\textbf{Assessing the impact of costly punishment and group size in collective-risk climate dilemmas}
}
\end{flushleft}

\bigskip

\begin{flushleft}
\bigskip

{{\bf Luo-Luo Jiang$^{1}$, Zhen Wang$^{2,3,4}$, Chang-Song Zhou$^{2,3}$,
Jurgen Kurths$^{5,6,7}$, and Yamir Moreno$^{8,9,10}$}

\bigskip
\vspace{5ex}
$^{1}$ College of Physics and Electronic Information Engineering, Wenzhou University, 325035 Wenzhou, China\\
$^{2}$ Northwestern Polytechnical University, Xian 710071, China\\
$^{3}$ Department of Physics, Hong Kong Baptist University, Kowloon Tong, Hong Kong\\
$^{4}$ Center for Nonlinear Studies and the Beijing-Hong Kong-Singapore Joint Center for
Nonlinear and Complex Systems (Hong Kong), Hong Kong Baptist University, Kowloon Tong, Hong Kong\\
$^{5}$ Potsdam Institute for Climate Impact Research (PIK), 14473 Potsdam, Germany\\
$^{6}$ Department of Physics, Humboldt University, 12489 Berlin, Germany\\
$^{7}$ Institute for Complex Systems and Mathematical Biology, University of Aberdeen, Aberdeen AB24 3UE, United Kingdom\\
$^{8}$ Institute for Biocomputation and Physics of Complex Systems (BIFI), University of Zaragoza, Zaragoza 50009, Spain\\
$^{9}$ Department of Theoretical Physics, University of Zaragoza, 50009 Zaragoza, Spain\\
$^{10}$ISI Foundation, Turin, Italy.
%Electronic address:
%$^{\ast}$\url{zhenwang0@gmail.com};$^{\dagger}$\url{Juergen.Kurths@pik-potsdam.de}
}

\end{flushleft}

\bigskip
\vspace{22ex}

\noindent
\textbf{The mitigation of the effects of climate change on humankind is one of the most pressing and important collective governance problems nowadays $^{\textbf{1-4}}$. To explore different solutions and scenarios, previous works have framed this problem into a Public Goods Game (PGG), where a dilemma between short-term interests and long-term sustainability arises $^{\textbf{5-9}}$. In such a context, subjects are placed in groups and play a PGG with the aim of avoiding dangerous climate change impact. Here we report on a lab experiment designed to explore two important ingredients: costly punishment to free-riders and group size. Our results show that for high punishment risk, more groups succeed in achieving the global target, this finding being robust against group size. Interestingly enough, we also find a non-trivial effect of the size of the groups: the larger the size of the groups facing the dilemmas, the higher the punishment risk should be to achieve the desired goal. Overall, the results of the present study shed more light into possible deterrent effects of plausible measures that can be put in place when negotiating climate treaties and provide more hints regarding climate-related policies and strategies.}
\vspace{10ex}

As it is well-known, human activities have already demonstrably changed global climate through the emission of greenhouse gases, specially $CO_2$, into the atmosphere \cite{o2002climate,manabe1993century}. If we do not reduce the release of these greenhouse gases, much greater changes, such as global warming and sea-level rise, will become inevitable consequences \cite{broecker1997thermohaline,hansen2005slippery}. As a matter of fact, the Intergovernmental Panel on Climate Change (IPCC) is pushing for a reduction in greenhouse gas emission of about $50\%$ of the current level by 2050 \cite{tongji,peters2013challenge}. It is clear that achieving this target is a major collective governance problem, that requires coordination and cooperation of different countries and regions. A dilemma thus naturally poses itself: a severe reduction might depress economy and lead to less short-term economic benefits, whereas implementation of insufficient -or no- measures might cause severe climate changes and huge economic losses in the mid to the long term.

To mimic this type of social dilemmas, a number of game models, stylizing the climate change problem with countries or governments as players, have been proposed during the past years \cite{regan2016,bechtel2013mass,dietz2011paths}. Typical examples include: threshold public goods games, requiring a minimal investment into a common pool \cite{Milinski_pnas06}; emission games, where each actor can only release a certain amount of $CO_2$ per year \cite{svirezhev1999emission}; climate negotiation games, which need a special negotiation scenario \cite{barrett2012climate}; dynamic climate-change games, involving stochasticity and scientific uncertainty \cite{dutta2004self} and collective-risk social dilemmas, where the investment aims to avert the risk of losing more benefits due to climate change \cite{Milinski_pnas08}. Among these existing frameworks, collective-risk social dilemmas have attracted most attention, both theoretically and experimentally \cite{Vasconcelos_n13,chen2012risk,hilbe2013evolution}. In this simple, paradigmatic setup, subjects are divided into groups and repeatedly make decisions of investment with a target goal in mind, that represents the minimum amount the group needs to invest to avoid the undesired outcome. In the present context, achieving the goal means that dangerous climate change impact could be mitigated, otherwise the remaining individual wealth is at stake and can be completely lost with a certain (loss) probability.

A recent collective-risk climate dilemma research \cite{Milinski_pnas08} found particularly interesting results: the higher the risk of losing the accumulated earnings is, the easier it is to reach the collective target sum. In other words, climate change mitigation is more likely to be achieved when the probability of mid- and long-term climate impact is higher. However, despite the previous insights, there is a relevant scenario that remains largely unexplored: the effects of punishment, which penalizes free-riders -or non-cooperative individuals \cite{hauert_s07,Milinski_p06,barrett2002towards}. Importantly enough, such a strategic change in collective dilemmas can be mapped to actual policies as recently discussed \cite{regan2016,barrett2008climate,barrett2010contrasting}.

The question, thus, becomes how costly punishment influences individual investment in the collective-risk climate dilemma game. Additionally, we also study here the influence of different group sizes, as this might also have implications in current negotiations to implement measures to mitigate global climate change impact. Our findings show that both ingredients play a key role in reaching the final goal: while costly punishment increases the likelihood to collect the desired amount, the dependence of the group size is less intuitive. We uncover that the larger the group size is, the harder it is to accomplish the collective target for even large values of the punishment risk. These results point to the existence of a non-trivial tradeoff between enforcing measures and cooperative multi-country governance in climate treaties.

To explore the aforementioned issues, we have carried out a lab experiment where subjects were divided into independent groups of size $M$. Initially, all individuals had 20 monetary units (MU), and each subject was able to contribute 0, 1 or 2 MU to her group. If after 10 rounds the collective target of $10\cdot M$ was achieved, players keep the money they saved. At variance with the traditional setup \cite{Milinski_pnas08,jacquet2013intra}, we introduced costly punishment: if there are free-riders -individuals that invest 0 MU-, with a probability $p$, referred to henceforth as punishment risk, such subjects are fined with 3 MU. Moreover, punishment is not cost-free but has a cost of 1 MU that is evenly shared by all group members. Note that in this way, whether to punish free-riders or not  is not decided by the rest of the players in the game; instead, it is applied with certain probability. Finally, if the group does not reach the target amount, with probability 0.5, all the individuals lose their savings, otherwise the remaining amount constitutes their earnings. See Methods and Supplementary Material (SM) for further details on the experimental design.

\begin{figure}
\renewcommand{\figurename}{\textbf{Figure}}
\scalebox{0.65}{\includegraphics{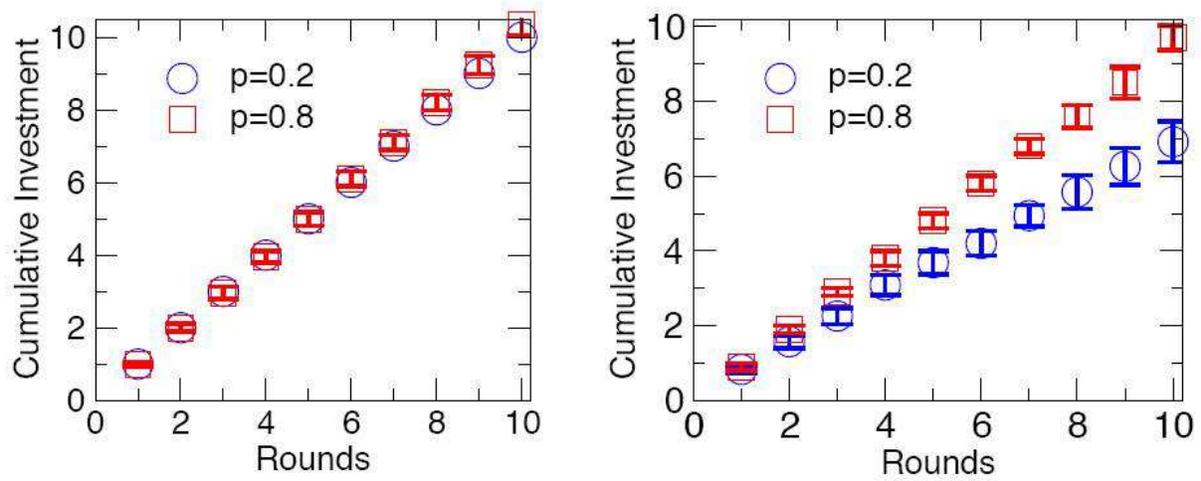}}
\caption{\label{fig:evolution}\textbf{Evolution of the cumulative investment with the number of rounds.} The two panels show the cumulative amounts contributed to the PGG as a function of the number of rounds played for groups of size $M=5$ and two values of the punishment risk $p$. The left panel displays how this quantity varies when considering only the games in which the final target was achieved, whereas the right panel shows results averaged over games in which it was not. Interestingly, even in the case in which the target was almost unreachable (left panel, $p=0.2$), the players kept donating, which is a consequence of the punishment mechanism and that the loss probability was set to 0.5. Error bars represent the Standard Error of the Mean (SEM).}
\end{figure}

We first show in Figure~\ref{fig:evolution} results obtained for the evolution of the collected amount in both games in which the final target was achieved (left panel) and was not (right panel), as a function of the punishment risk $p$ for groups of size $M=5$. As can be seen in the figure, those games in which the final amount required was reached are dominated by a steady increase in the cumulative investments, without abrupt changes in the shape of the curves, for both values of $p$. In fact, the slope is roughly one, indicating that on average, individuals contributed 1MU per player in each round of the game. The behavior for the cases in which the final goal was not attained is however dependent on $p$. As the punishment risk increases, the average amount invested per round is higher. Interestingly enough, even for high values of $p$, it is most of the times not enough. However, players do not stop contributing to the PGG, though they invest less and less as they approach the last round. Even if the probability of losing everything left at the end of 10 rounds is $1/2$, the latter behavior is rooted in the need to avoid additional loses that players might incur in if they act as free-riders: as fines are imposed with high probability, any eventual savings might be taken out by the fine itself or by the cost of applying it after 10 rounds.

\begin{figure}
\renewcommand{\figurename}{\textbf{Figure}}
\scalebox{0.65}{\includegraphics{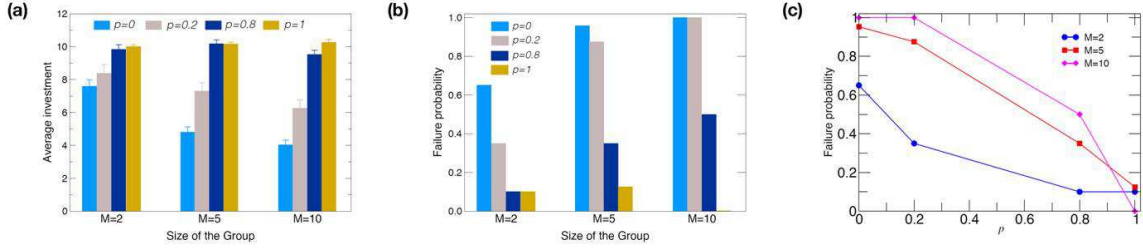}}
\caption{\label{fig:groups} \textbf{Average investment and failure probability as a function of the group size $M$ and $p$.}  Panels (a) and (b) shows both quantities for four different values of the punishment risk $p$, while panel (c) represents the fraction of failures as a function of $p$ and $M$. In (a) and (b), and for all groups of histograms, the values of the punishment risk $p$ are as indicated. We found that punishment is effective only for high values of $p$, meaning that free-riders will be fined with a high probability. The dependence with the group size shows that for all values of $p<1$, the larger the size of the group is, the less amount is contributed and the harder it is to achieve the final goal. Error bars represent the Standard Error of the Mean (SEM).}
\end{figure}

Figure~\ref{fig:groups} shows the cumulative investments averaged over all the members of the group and the failure probability, as a function of the size of the groups for four different values of the punishment risk $p$ after 10 rounds. We also show in panel (c) the same results displayed in (b), but represented as a function of $p$ for fixed values of the group sizes. The left (light blue) bars in each set of Figures.~\ref{fig:groups}(a) and (b) display results for $p=0$, that corresponds to the situation in which there is no punishment to free-riders. Two features are worth highlighting: the number of experiments for which the target amount was not achieved (failure probability) is remarkably high, which in turn increases with the size of the group, see also Figure~\ref{fig:groups}c.  This is a consequence of the low amount contributed to the PGG in all cases,  $7.5785\pm0.39002$, $4.79167\pm0.32773$ and $4.025\pm0.28662$ for $M=2$, $M=5$ and $M=10$, respectively. However, when the punishment (namely, $p>0$) comes into play, the fraction of failures starts to drop, leading to an increase in the number of PGG in which the final target is reached. Interestingly, this decrease of the fraction of failures is significant only when the probability of being fined is large enough. As for the dependence with the size of the group, the same pattern with respect to the case $p=0$ is observed, as shown more explicitly in Figure~\ref{fig:groups}c. Indeed, only when $p=1$, the probability of failure is measured to be 0 for the largest group in our experiments ($M=10$) and the average amount contributed increases with $M$ ($10\pm0.12403$, $10.15\pm0.1163$ and $10.25\pm0.17813$ for $M=2$, $M=5$ and $M=10$, respectively). We show in the SM that these results are robust against the loss probability (Figure S5). These findings hence suggest that the larger the size of the groups is, the higher the punishment to free riders should be for the final goal to be attained.

\begin{figure}
\renewcommand{\figurename}{\textbf{Figure}}
\scalebox{0.58}{\includegraphics{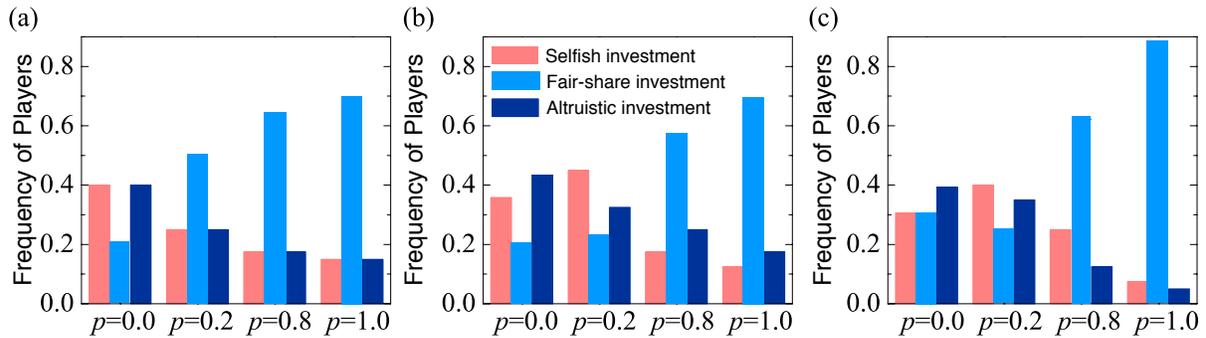}}
\caption{\label{fig:investment} \textbf{The distribution of classes of players as a function of the punishment risk $p$ for different group sizes $M$.} Increasing the punishment likelihood reduces the number of free-riders in the game, but at the expense of a decrease in the number of maximal contributors. As a result, most of the players adopt a fair-share strategy, which however is not enough to reach the final target in many of the games, as indicated by the fraction of failures in Fig\ \ref{fig:groups}. The sizes of the groups are, from (a) to (c), $M=2$, $M=5$ and $M=10$, respectively.}
\end{figure}

Next, we analyze individual behavior. We consider three different possibilities: i) selfishness, typifying free riders that contribute 0 to the PGG but obtain the largest benefit if the global target is achieved; ii) fairness, characterizing those individuals contributing the fair-share of 1MU; and iii) altruism, describing the behavior of those individuals that contribute the most (2MU). Figure~\ref{fig:investment} shows the distribution of the three behaviors as a function of the punishment risk $p$ for different group sizes. As it can be seen, regardless of the group size, the number of free-riders decreases in general when the punishment risk is nonzero and grows. The opposite trend is observed concerning the number of subjects contributing the fair-share. Interestingly enough, the deterrent effect of punishment makes altruistic contributions to decrease as well. Admittedly, the selfish and altruistic investments go hand-to-hand, namely, as soon as the number of free-riders decreases due to the higher values of the punishment likelihood, the number of maximal contributions does not remain constant but also decreases in favor of the fair-share behavior. This might indicate that players realize that a fair-share, which is the least amount an individual can contribute without incurring in a fine, is enough to reach the target while maximizing final benefits. Indeed, as seen in Figure~\ref{fig:evolution} for the games in which the final goal was achieved, there is no abrupt change in the amounts contributed in each round of those games, which is a further indication that the fair-share strategy is quite stable as rounds go by. As for the dependence with the group size, we do not observe significant variations of the previous patterns, except for the largest size $M=10$ and $p=1$, a scenario in which almost all players ($87\%$) contribute the fair amount. Note that the latter case shows the lowest number of free-riders but also of altruism level.

Finally, we have also explored the relation between the punishment risk and both the average investment and the obtained payoffs. Results for $M=5$ are shown in Figure~\ref{fig:payoff}(a), where it can be seen that as $p$ increases, the average contribution to the common pool also grows. Moreover, as more groups achieve the final target, also the players' payoffs increases with the probability of being fined. Note that, although the average contributed amount increases significantly as soon as $p>0$, it is still far from the average amount needed to reach the final target. For low values of $p$, this persistent noncooperative behavior might be due to the fact that the actual risk of being punished is not very high. Indeed, for $p\le \frac{M(1-p*)}{3M+1}$ (where $p*$ is the loss probability, see Methods), the more rational strategy to maximize benefits is to free ride. As $p$ increases beyond the previous bound, adopting a fair-share strategy is the best, as even a single defector would earn less. However, as it can be seen in the figure, only for high values of $p$ (beyond $p=0.8$ in our case), punishment has an impact on the players' behavior. To check what is the perceived risk of being punished with respect to the actual value of $p$, we also asked to the subjects, every time they acted as free-riders, whether they thought they will incur in a fine (see Figure S4 of the {\em SM}). Figure~\ref{fig:payoff}(b) (see also Table~\ref{tab:deterrence2}) shows the variation of the estimated punishment ratio, $q$, measured as the ratio between the number of times subjects believed they will be punished and the number of times they opted to play as free-riders, as a function of $p$. As can be clearly seen, the perceived risk is always below the diagonal, i.e., that most of the players are not risk-averse, that is, most of the times that they did not contribute they were willing to take the risk -conjecturing that they would not be fined. The observed behavior in its turn also explains why punishment is effective only when it takes place with high probability.

Summarizing, the results of the present collective-risk social dilemmas experiments have important implications. Even if our experimental setup does not capture all the complexity of a collective governance problem such as agreeing on measures to mitigate dangerous climate changes, it certainly gives further insights into a class of dilemmas -the tragedy of the commons \cite{hardin1968tragedy}- that can provide hints to interpret and shape the dynamics of climate summits. Firstly, our findings show that punishment could be an effective mechanism to achieve global targets in the current context. At the same time, however, we have shown that in order for such an enforcing measure to be efficacious, it should be perceived as almost certain, otherwise its effects might be blurred. Hence, if we realize that individuals, institutions or the private sector are hesitant to provide a collective good without being enforced because the short-term benefits of defection are higher, then it follows that international treaties should necessarily compel governments to adopt measures aimed at overcoming those short-term incentives to free ride.

Secondly, our results regarding the dependence on the group size - an aspect that has gone largely unexplored before- are potentially far reaching: increasing the size of the group does not make the final goal more feasible, on the contrary, the fraction of failures might increase unless the punishment is almost always imposed. Our findings also provide an experimental confirmation of Olson's theory \cite{regan2016}, that relates both the composition and size of a group to its ability to achieve some socially optimal collective good, like sustainable growth without dangerous climate changes. Specifically, Olson argued that the larger the group size, the easier that a defection goes unnoticed by the rest of the group members, which ultimately leads to a number of free riders in the group that impedes to reach the collective good. Importantly, we have shown that the same arguments hold when free-riders (and to some extent the whole group as each fine carries a cost) are punished with certain probability.

In conclusion, and with all due caution, the present study suggests that in climate negotiations, measures such as imposing economic sanctions to non-cooperative countries and reducing the size of the groups negotiating treaties might be effective. Finally, we also stress that the current experiment has been done with a pool of subjects from China. Taking into account that this sort of collective-risk games is typically conducted in Western countries, we might expect that differences arise due to geoeconomics and cultural factors. However, compared to the previous result available \cite{Milinski_pnas08}, our results are not very different. Specifically, Ref.\cite{Milinski_pnas08} reported, for groups of size six, that with a loss probability of $1/2$ only 1 out of 10 groups reached the target. That is the same result obtained in our experiment for $M=5$ and $p=0$, the closest comparable setup. The latter similarity between both studies deserves to be further explored in future experiments, since any climate treaty negotiation is multilateral by nature, and thus, it would be important to replicate this and similar experiments involving subjects from different countries.

\vspace{5ex}
\begin{Large}
\noindent
\textbf{Methods}
\end{Large}
\vspace{5ex}

\noindent
\textbf{Maximizing payoffs.} As mentioned before, even with punishment in place, free-riding could be the more rational strategy to maximize benefits depending on the value of the punishment risk $p$ and of the loss probability $p^*$. To see this, let us assume that all players behave in the same way. In one scenario, where all players are free-riders, the target will never be reached. In the second scenario, where all players adopt a fair-share strategy, the target will always be attained. Thus, when is it more profitable in terms of the likelihood to maximize benefits to play as a free-rider or as a fair-sharer? The final expected payoff of the free-riders in the first situation would be:
\[
\Pi^{f}_{(1)}=\Pi^{i}-\sum_{j=1}^{N}(3p+\frac{p}{M}),
\]
while in scenario two it would be:
\[
\Pi^{f}_{(2)}=\Pi^{i}-\sum_{j=1}^{N}x,
\]
where $\Pi^{f}_{(\cdot)}$ and $\Pi^{i}$ are, respectively, the final expected payoff and the initial capital, $M$ is the size of the group, $N$ is the number of rounds played, and $x$ is the contribution to the common pool. Thus, for defection to be better than fair-share, the relation $\Pi^{f}_{(1)}\ge\Pi^{f}_{(2)}$ should be verified, which leads to the condition (setting $x=1$):
\[
p\le \frac{M}{3M+1},
\]
As free-riders only collect their benefits with probability $(1-p*)$, the final condition for $p$ is
\[
p\le \frac{M(1-p*)}{3M+1}.
\]

\noindent
\textbf{Experimental sessions.} The experiments of collective-risk social dilemmas were conducted from July of 2013 to November of 2013 and they involved a total of 720 freshmen and sophomore (coming from different majors, see Section 1.1 in {\em SM}) at Wenzhou University, China. Subjects consent was obtained before starting the experiments and after they answered a questionnaire (see Figure S1 in {\em SM}).
Each experimental session required the simultaneous participation of 20 subjects (see {\em SM} for more details), who were randomly divided into several groups (namely, the setup was completely anonymous). Both the size $M$ and composition of the groups were kept constant during the whole experiments. Within each group, subjects repeatedly played 10 independent rounds of the game. Each subject started with an initial endowment of 20 monetary units (MU). Each experimental round consisted of the following steps:
\begin{itemize}
\item At each round, all the subjects were asked simultaneously whether they would independently contribute 0 MU, 1 MU, or 2 MU to the climate account.
\item After taking their investment decisions, every subject was shown the following information during 30 seconds: ($i$) individual contribution (0 MU, 1 MU, or 2 MU); ($ii$) the collective investment of his/her group in the current round; ($iii$) the remaining gap between the cumulative contribution and the required target sum of the group (see Section 1.2 in {\em SM}).
\end{itemize}

Similar to previous experimental setups \cite{Milinski_pnas08}, the total investment required for one group of size $M$ to reach its target was set to $10\cdot M$ (equivalent to 1 MU per subject per round on average). If the overall contribution after 10 rounds was equal or greater than the collective target, individuals could keep the money saved. On the contrary, if the target sum was not reached, subjects could lose all their savings with a (loss) probability that we set to 0.5 in most of the sessions carried out.

Based on the above-mentioned basic setup of the collective-risk dilemmas game, we introduced costly punishment into the experiments. Distinguishing from previous theoretical researches about pool- or peer-punishment in game theory \cite{Milinski_p06}, the implementation of punishment in the present work is directly related to the performance of the group, and the cost of imposing a fine to non-cooperators was evenly distributed among all group members, regardless of their investment behavior. More specifically, if at any round of the game there exist non-cooperators -people that contribute zero to the PGG- a fine of 3 MU is imposed to those selfish players with probability $p$. At the same time, the total cost of punishment, namely, 1MU per selfish player -since the cost of 1MU is associated to each fine applied- is equally distributed among the $M$ players of the group.

Finally, at the end of the experimental session, the remaining monetary units (MU) were changed into real money. Earnings -including the show-up fee- ranged from 20$\yen$ to 40$\yen$ and the conversion rate applied was 1 MU = 1$\yen$. Altogether, the results reported here come from 232 groups (120 groups of size $M=2$, 64 of size $M=5$ and 16 of size $M=10$). The instructions and questionnaires took 5 to 10 minutes and the entire game took 30 to 35 minutes for the 10 rounds. The average earning of all the participants was 32.2 $\yen$.

\vspace{5ex}
\noindent
\textbf{Acknowledgments}~~We would like to thank K.-Z. Jin, C. Gracia-L\'azaro and A. S\'anchez for helpful discussions. This work was supported by the National Natural Science Foundation of China (Grants 61203145, 11047012 and 91024026).

\vspace{2ex}
\noindent
\textbf{Competing financial interests}~~The authors declare that they have no competing financial interests.

\vspace{2ex}
\noindent
\textbf{Author contributions}~~LLJ, ZW designed the experiments. LLJ, ZW analyzed the data. All authors contributed to the analysis of the results and edited the manuscript. LLJ, ZW and YM were the leading writers of the manuscript.

\newpage
\clearpage

%\begin{table}
%{\small
%\begin{center}
%\caption{\label{strategy} \textbf{The classification of subjects' investment behaviors according to the requirement of collective target.}}
%\begin{tabular*}{0.95\columnwidth}{@{\extracolsep{\fill}} |c|c|c|c|}
%%\begin{tabular*}{0.95\columnwidth}{|c|c|c|c|}
%%\hline
%\hline
%\bf Investment behavior    &\tabincell{c}{\bf Contribution\\ \bf per round} & \tabincell{c}{\bf Minimum Investment per \\ \bf subject per round on average} &\bf Collective target* \\
%\hline
%\small Selfish investment      & \small 0     &   &  \\ \cline{1-2}
%\small Fair-share investment   & \small 1     & \small 1  & \small $10M$\\ \cline{1-2}
%\small Altruistic investment   & \small 2     &    &  \\
%\hline
%%\hline
%%\end{tabular*}
%
%\begin{tablenotes}
% *Here the collective target $10\cdot M$ (MU) is the minimal required sum, to avert the risk of losing more benefits due to dangerous climate change, in 10 repeated interactions with the groups of size $M$. Thus, the basic condition for reaching such a goal is equivalent to 1 MU per subject per round on average.
%\end{tablenotes}
%%\end{threeparttable}
%\end{center}
%}
%\end{table}

\begin{table}
\begin{center}
\caption{\label{tab:deterrence2} \textbf{Statistical results for number of actual
punishment events, estimated punishment events and non-investment cases with different
punishment risk $p$.}}
%\begin{tabular}{7cm}{@{\extracolsep{\fill}}l | c c c}
\begin{tabular*}{0.8\columnwidth}{@{\extracolsep{\fill}} |c | c| c| c| c| c|}
\hline
\hline
\bf  Punishment risk ($p$)    &\small 0.2 &\small 0.4 &\small 0.6 &\small 0.8 &\small 1.0  \\
\hline
\bf Num. of actual punishment events   & \small 29  &\small 34  &\small 43 &\small 16  &\small 4\\
\hline
\bf Num. of estimated punishment events   & \small 16  &\small 21  &\small 19 &\small 6  &\small 3\\
\hline
\bf Num. of non-investment cases   & \small 126  &\small 97  &\small 86 &\small 18  &\small 4\\
\hline
\hline

\end{tabular*}
\end{center}
\end{table}

\clearpage

\noindent  \textbf{\Large Figures Captions.} \\

\clearpage

\begin{figure}
\renewcommand{\figurename}{\textbf{Figure}}
\scalebox{0.6}{\includegraphics{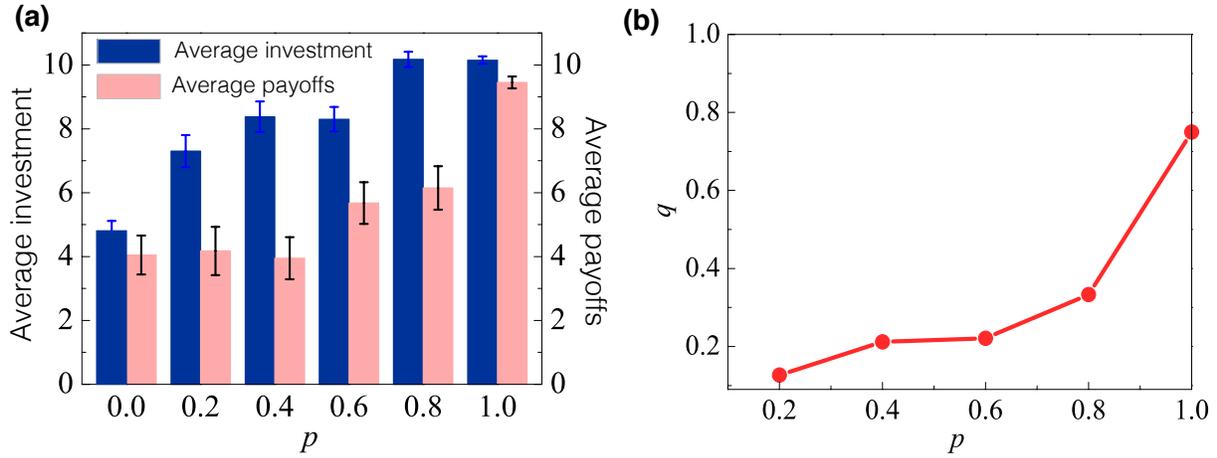}}
\caption{\label{fig:payoff} \textbf{Dependence of the average investments, payoffs and risk-perception on $p$}. Panel (a) shows how the average investments and the payoffs vary with the likelihood of punishment $p$ for groups of size $M=5$. As observed, only high values of $p$ induce players to contribute what is needed on average to reach the final target. Interestingly, this might be due to the perception that the risk to be punished, $q$, is low, as shown in panel (b). This latter quantity is measured as the ratio between the number of times a player thought she was going to be punished and the number of times the same subject played as free rider. Error bars represent the Standard Error of the Mean (SEM).}
\end{figure}

\clearpage
\begin{flushleft}
{\Large
\textbf{Assessing the impact of costly punishment and group size in collective-risk climate dilemmas}
}

\bigskip

{\Large
\textbf{Supplementary Information}
}

%\bigskip

\bigskip

{\large {\bf Luo-Luo Jiang$^{1}$, Zhen Wang$^{2,3,4}$, Chang-Song Zhou$^{2,3}$,
Jurgen Kurths$^{5,6,7}$, and Yamir Moreno$^{8,9,10}$}

\bigskip

$^{1}$ College of Physics and Electronic Information Engineering, Wenzhou University, 325035 Wenzhou, China\\
$^{2}$ Northwestern Polytechnical University, Xian 710071, China\\
$^{3}$ Department of Physics, Hong Kong Baptist University, Kowloon Tong, Hong Kong\\
$^{4}$ Center for Nonlinear Studies and the Beijing-Hong Kong-Singapore Joint Center for
Nonlinear and Complex Systems (Hong Kong), Hong Kong Baptist University, Kowloon Tong, Hong Kong\\
$^{5}$ Potsdam Institute for Climate Impact Research (PIK), 14473 Potsdam, Germany\\
$^{6}$ Department of Physics, Humboldt University, 12489 Berlin, Germany\\
$^{7}$ Institute for Complex Systems and Mathematical Biology, University of Aberdeen, Aberdeen AB24 3UE, United Kingdom\\
$^{8}$ Institute for Biocomputation and Physics of Complex Systems (BIFI), University of Zaragoza, Zaragoza 50009, Spain\\
$^{9}$ Department of Theoretical Physics, University of Zaragoza, 50009 Zaragoza, Spain\\
$^{10}$ISI Foundation, Turin, Italy.
%Electronic address:
%$^{\ast}$\url{zhenwang0@gmail.com};$^{\dagger}$\url{Juergen.Kurths@pik-potsdam.de}
}

\end{flushleft}

\bigskip

\noindent \textbf{\large S1. Experimental Materials and Methods}\\

\bigskip

\noindent \textbf{\large 1.1 Volunteer Recruitment and Treatment}\\

After the call of participation, in which no details about the experiment to be conducted were given, volunteers were required to show up in the experimental labs the days of the experimental sessions. Participants were divided into 36 sessions that were assigned at random. In each session, every participant was assigned to one isolated computer cubicle, where mutual communication was forbidden. When participants sat down, they were required to read the instructions, which explained the collective-risk social dilemma experiments, on the screen. After reading the tutorial, they were once again required to finish the questionnaires within a given time (the questionnaire is provided in Figure~\ref{FigS2}, whose translation to English is provided below). Only once the questions were answered accurately, the participants could engage in the experiments. Otherwise, the participant was not allowed to play the games and she received a show-up fee of 20$\yen$. The selection of participants ended when each section had 20 eligible subjects. Thus, there were 720 qualified volunteers in total that engaged in the experimental sessions, of which 398 were males and 322 were females representing 55.28\% and 44.72\% of the total number of eligible participants, respectively. As for the participants' background, they were students from the following major degrees: mathematics (55), physics (56), electronic information (68), chemistry (58), biology (50), computer science (60), business management (47), law (51), history (32), Chinese language culture (18), machinery engineering (60), economics (40), English (45), advertisement design (40) and civil engineering (40).

During the realization of the experiments, we let at least two teachers supervise each session (any interaction among the subjects was forbidden). Each session lasted about 30 minutes and at the end of the sessions, all the participants cashed their earnings, which ranged from 20$\yen$ to 40$\yen$, including the show-up fee.

\bigskip
\bigskip
\noindent
\textbf{Translation of the questionnaire in climate game experiments}

\begin{itemize}
\item[]{Case 1:}

During the collective-risk social dilemma game, each group includes 5 subjects, who respectively have 20 money units (MU) at the beginning and need to repeatedly play 10 independent rounds of the game. A total investment is required to reach the target of 50 MU, which can help to avoid the dangerous climate change. For example, given 12 MU as your total investment and other partners' contribution being 40 MU, the extreme climate will be well prevented (since 12+40=52$>$50). After the game, you can earn 20-12=8 MU. In addition, because you need to share a punishment cost of 0.8 MU during the game, your final earning will be (this has to be filled out by participants) {\bf 7.2} MU.

\item[]{Case 2:}

During the collective-risk social dilemma game, each group includes 5 subjects, who respectively have 20 money units (MU) at the beginning and need to repeatedly play 10 independent rounds of the game. A total investment is required to reach the target 50 MU, which can help to avoid the dangerous climate change. For example, given 8 MU as your total investment and other partners' contribution being 41 MU, extreme climate changes will happen with 50\% probability (since 8+41=49$<$50). If there is the extreme climate, all 5 members would not get any earnings. In the opposite case, you would earn 20-8=12MU after the game. Moreover, because you were punished two times (the fine of each punishment is 3 MU) and you need to share he punishment cost of 0.4 MU during the game, your final earning will be (this has to be filled out by participants) {\bf 5.6} MU in this scenario.
\end{itemize}

\bigskip

\noindent \textbf{\large 1.2  Experimental Platform and Interface}\\

The platforms and interfaces were designed based on Z-tree \cite{fischbacher2007z}, which provides an excellent visual framework for participants. Before the first round, the subjects were shown the following information: (1) initial endowment of 20 monetary units (MU); (2) the size $M$ of the group that he/she belonged to; (3) the collective target $10\cdot M$. Participants knew that only if the group reached or exceeded the target sum, the dangerous climate change could be successfully prevented.

During each round, subjects needed to make an independent investment decision from three available choices: 0 MU, 1 MU or 2 MU, to the climate account (see Figure~\ref{FigS3} for schematic illustration), in which cumulative contributions of the group were added. In addition to their investment choices, subjects could also see at each round of the game their remaining monetary units, the gap between the total investment and the required target as well as the value of the punishment risk. This information was displayed, as every other interface, during 30 seconds. After making their decisions, subjects needed to click the "OK" button. If any one could not play within the specified time frame, the system produced a warning on the top of the screen, which reminded them to make the decision as soon as possible. Only after all the subjects made their decisions and clicked "OK", the system moved to a new interface.

If there is at least one subject that contributed zero, a new interface appeared in the screen of that subject -contributors were not affected by this and kept their interfaces as they were. In the new interface, selfish subjects were asked whether or not they felt they will be punished with the given punishment risk (see Figure~\ref{FigS4}). If punishment was eventually carried out, each selfish subject would immediately lose 3 MU as a fine and the expense (1 MU) of such a punishment event was evenly shared among all the group members.

After the selfish players finished and pressed again the "OK" button, the computer screens of all the subjects simultaneously moved to a new -common for all- interface (see Figure~\ref{FigS5}), where additional information was provided: whether or not a punishment event took place, the collective investment of his/her group in the present round, the number of non-contributors in the group and the remaining monetary units after punishment (even if it did not happen). Finally, after everyone pressed the "OK" button in this interface, the system moved to the next round.

The previous steps were repeated in all experimental sessions 10 times, which is the number of rounds set up.

We also mention that if groups finished in advance, the subjects of those groups were instructed to remain quiet in their positions until all the groups finished. Then, teachers in each session printed a receipt with the money earned, which was cashed at the same place and day according to the rate 1 MU=1$\yen$.

\bigskip
\bigskip

\noindent \textbf{\large S2. Experimental Materials and Methods}\\

Results in the main text were obtained for a loss probability of 0.5. Here we report results for two additional values of the loss probability to show that the findings in the main text are qualitatively robust against variation of this parameter. Specifically, Figure~\ref{FigS6} shows the individual average investment and the fraction of failures of groups as a function of the punishment risk $p$ for three values of the loss probability, as indicated.

\clearpage

\noindent  \textbf{\Large Figures in Supplementary Information } \\

\clearpage

\begin{figure}
\renewcommand{\figurename}{\textbf{Figure}}
\renewcommand\thefigure{S\arabic{figure}}
\setcounter{figure}{0}
\centerline{\epsfig{file=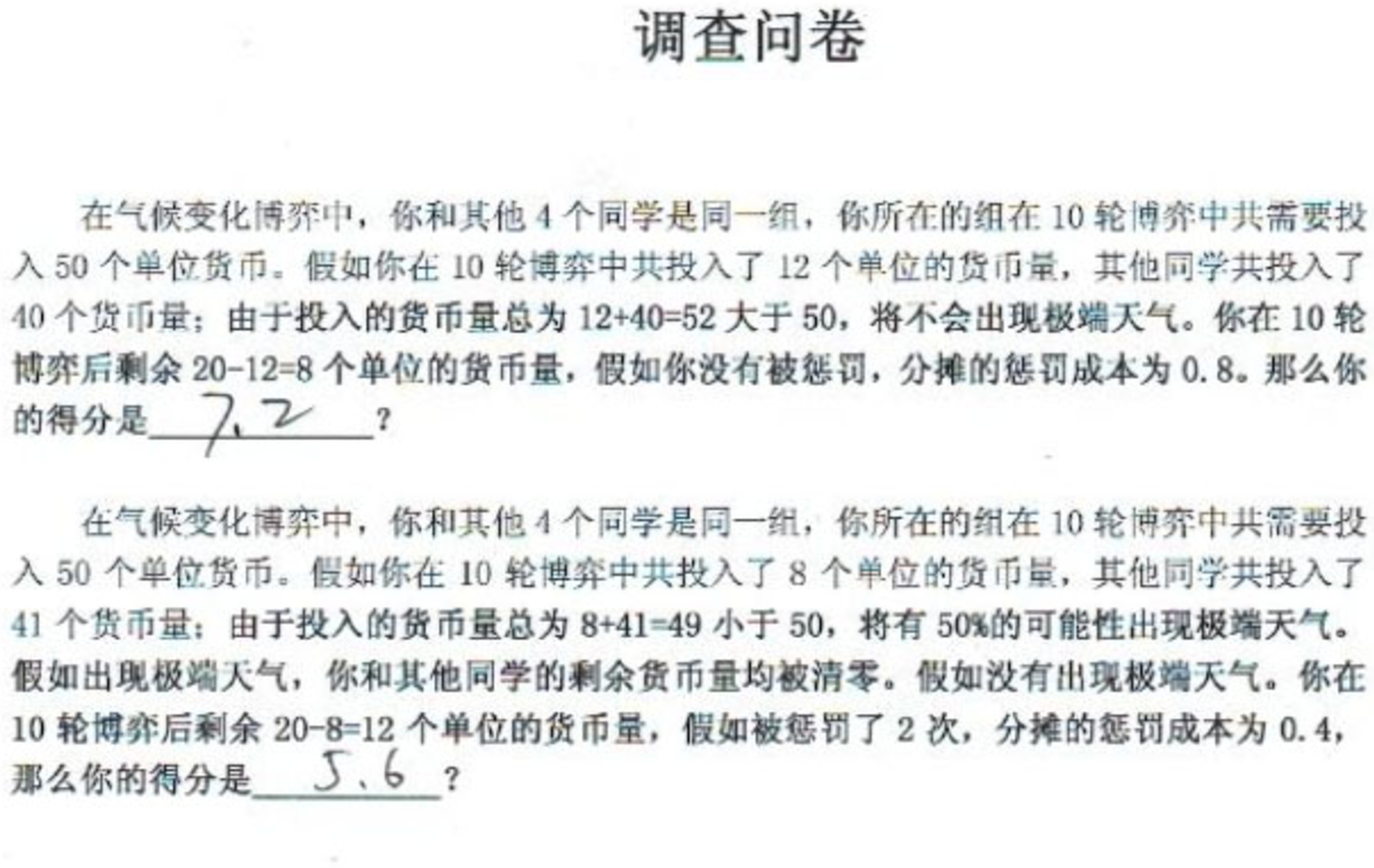,width=13cm}}
\caption{ \textbf{Snapshot of the questionnaire in the collective-risk social dilemma experiments with punishment.} The translation to English is provided in the SM text.}
\label{FigS2}
\end{figure}

\begin{figure}
\renewcommand{\figurename}{\textbf{Figure}}
\renewcommand\thefigure{S\arabic{figure}}
\centerline{\epsfig{file=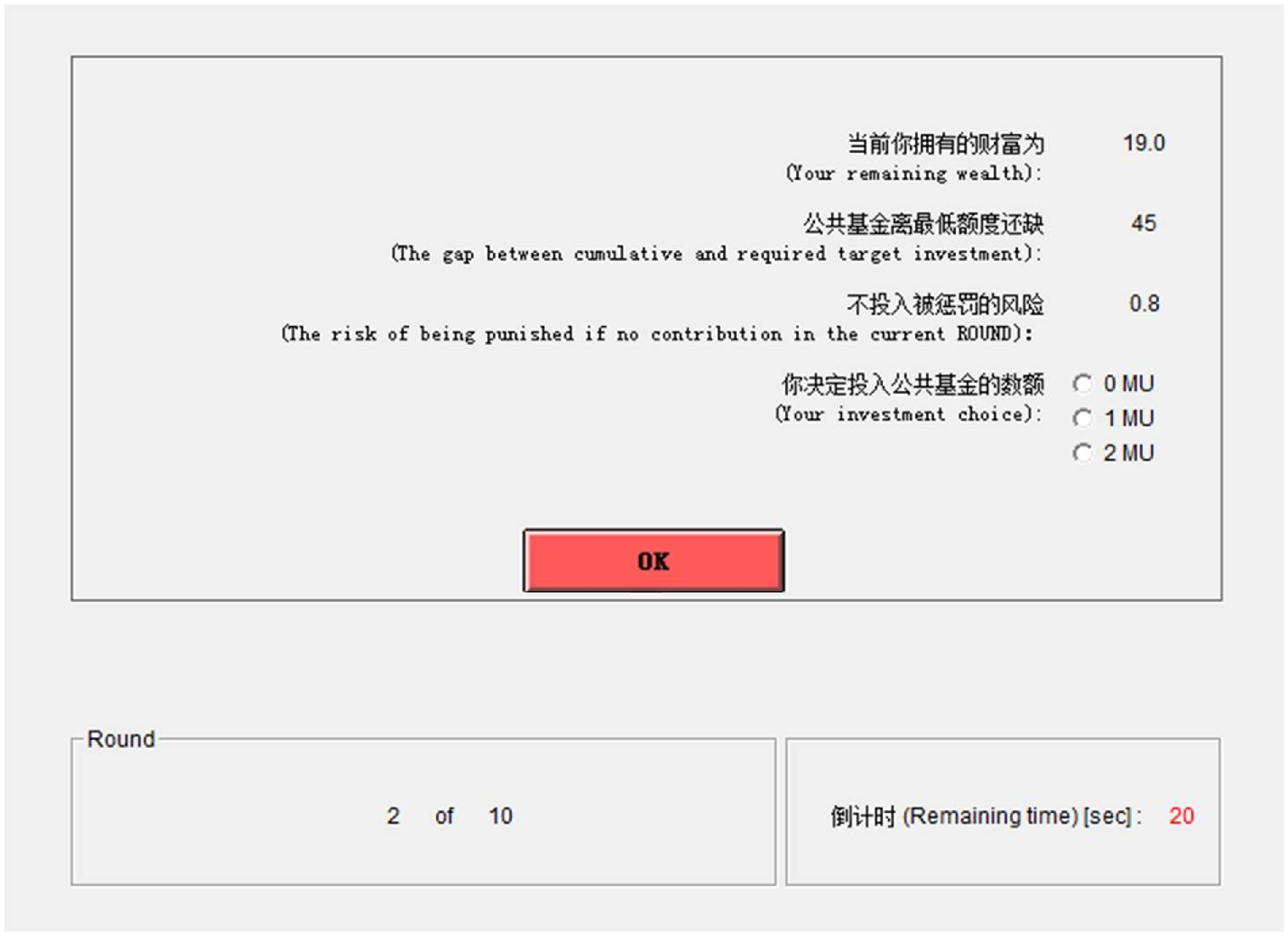,width=13cm}}
\caption{ \textbf{Interface of individual investment selection in each round. Besides, the additional information includes the number of round, remaining wealth of the subjects and the gap between cumulative investment and required target, and punishment risk.}}
\label{FigS3}
\end{figure}

\begin{figure}
\renewcommand{\figurename}{\textbf{Figure}}
\renewcommand\thefigure{S\arabic{figure}}
\centerline{\epsfig{file=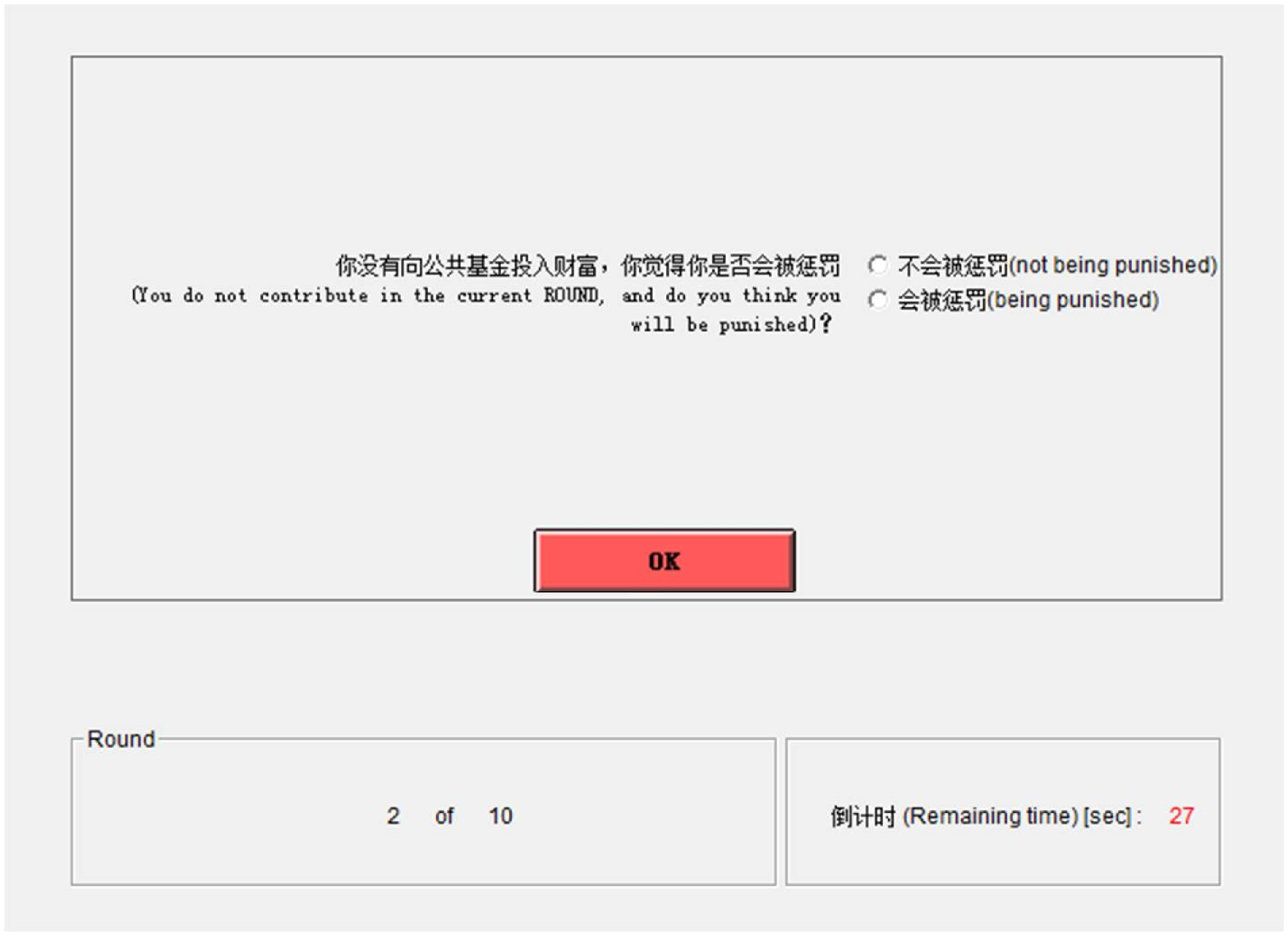,width=13cm}}
\caption{\textbf{Interface of estimating whether or not punishment event took place after the selection of selfish investment behavior.}}
\label{FigS4}
\end{figure}

\begin{figure}
\renewcommand{\figurename}{\textbf{Figure}}
\renewcommand\thefigure{S\arabic{figure}}
\centerline{\epsfig{file=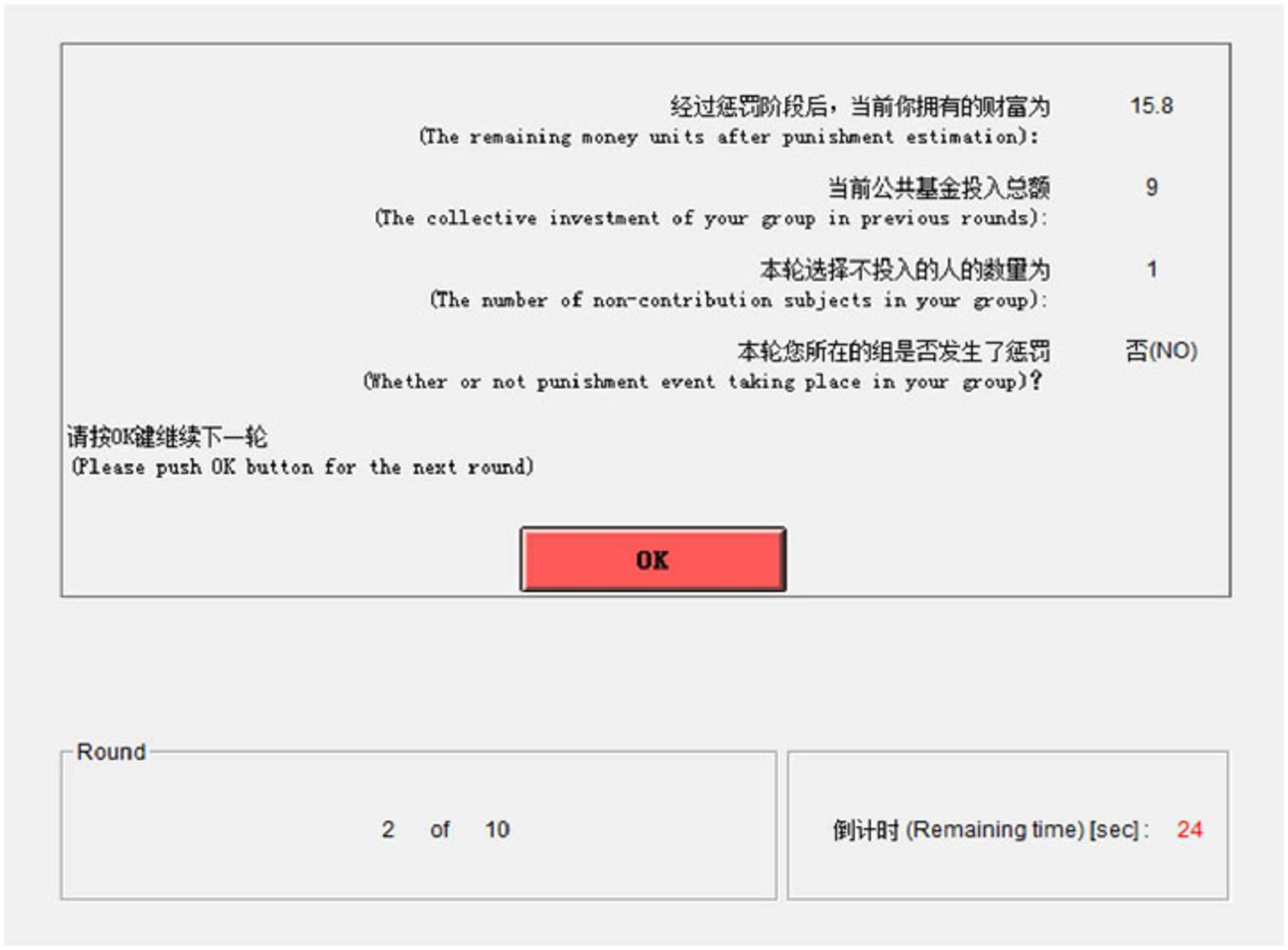,width=13cm}}
\caption{ \textbf{Interface of showing the remaining money units after punishment estimation. Besides, additional information also included whether or not punishment event taking place, collective investment of his/her group in the current round, the number of non-contribution objects in the group.}}
\label{FigS5}
\end{figure}

\begin{figure}
\renewcommand{\figurename}{\textbf{Figure}}
\renewcommand\thefigure{S\arabic{figure}}
\centerline{\epsfig{file=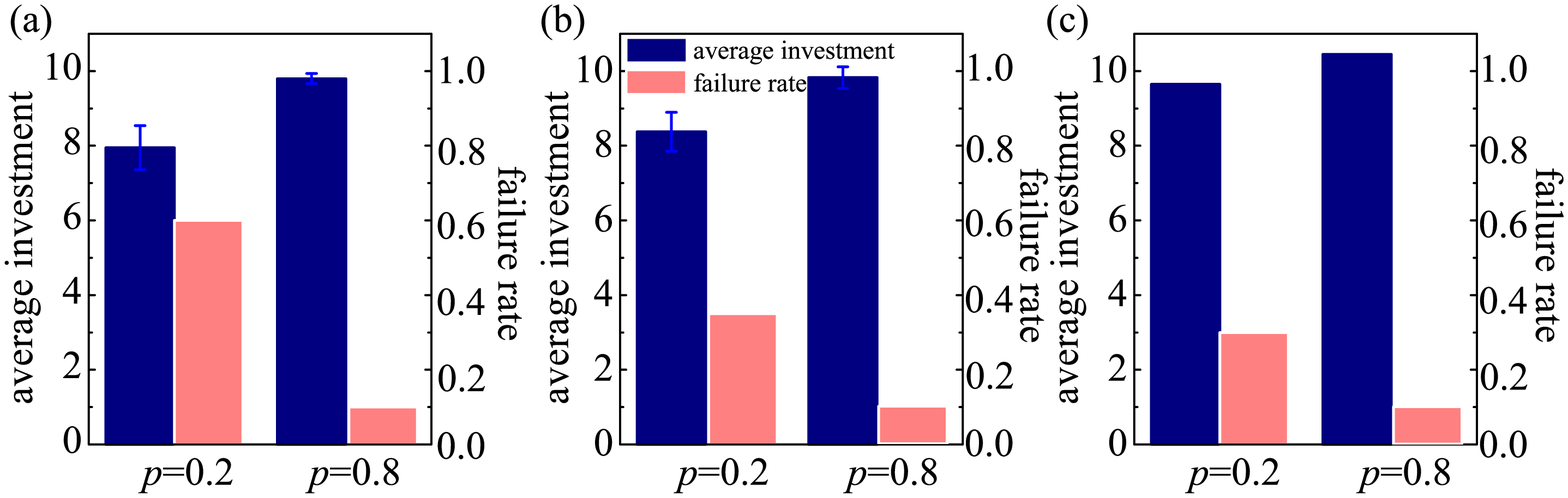,width=13cm}}
\caption{ \textbf{Average investment of subjects and failure rate of groups during 10 rounds of collective-risk dilemma game in dependence on punishment risk $p$ for different loss probability.} Here, the loss probability is the probability of losing reaming wealth once the collective investment does not reach the target. The values of failure rate are in the form of $mean \pm~s.e.m.$ From (a) to (c), the values of loss probability risk are 0.2, 0.5 and 0.8, respectively. The statistical quantities of Kruskal-wallis test are $\chi^2=7.95$ $d.f.=0.5915$ for $p=0.2$ and $\chi^2=9.8$ $d.f.=0.13765$ for $p=0.8$ (a); $\chi^2=8.375$ $d.f.=0.52218$ for $p=0.2$ and $\chi^2=9.825$ $d.f.=0.29284$ for $p=0.8$ (b); $\chi^2=9.65$ $d.f.=0.66203$ for $p=0.2$ and $\chi^2=10.45$ $d.f.=0.32827$ for $p=0.8$ (c).}
\label{FigS6}
\end{figure}

\end{document}